\documentclass[a4paper,fleqn,usenatbib]{mnras}
\usepackage{amsmath}
\usepackage{epsfig} % to include eps files - for texmaker
\usepackage{graphicx}	% Including figure files
\usepackage{amssymb}	% Extra maths symbols

\newcommand{\be}{\begin{equation}}
\newcommand{\e}{\end{equation}}
\newcommand{\bear}{\begin{eqnarray}}
\newcommand{\ear}{\end{eqnarray}}

\newcommand{\hmpc}{{\, h^{-1}\, {\rm Mpc}}}

\def\aj{AJ}
\def\apj{ApJ}
\def\apjs{ApJS}
\def\jcap{JCAP}
\def\mnras{MNRAS}
\def\aap{A\&A}
\def\prd{Physical Review D}
\def\nat{Nature}      
\def\apjs{ApJS}
\def\apjl{ApJ Letters}
\def\physrep {Physics Reports}

\title[Homogeneity on the largest accessible scale] {An information
  theory based search for homogeneity on the largest accessible scale}
    
\author[Sarkar, S. and Pandey, B.]  { Suman
  Sarkar\thanks{E-mail:sumansarkar.rs@visva-bharati.ac.in} and
  Biswajit Pandey\thanks{E-mail: biswap@visva-bharati.ac.in}
  \\ Department of Physics, Visva-Bharati University, Santiniketan,
  Birbhum, 731235, India\\ }
   
 \date{\today}

 \pubyear{2016}
  
\begin{document}
\label{firstpage}
\pagerange{\pageref{firstpage}--\pageref{lastpage}}      
\maketitle
       
\begin{abstract}
We analyze the SDSS DR12 quasar catalogue to test the large-scale
smoothness in the quasar distribution. We quantify the degree of
inhomogeneity in the quasar distribution using information theory
based measures and find that the degree of inhomogeneity diminishes
with increasing length scales which finally reach a plateau at $\sim
250 \hmpc$. The residual inhomogeneity at the plateau is consistent
with that expected for a Poisson point process. Our results indicate
that the quasar distribution is homogeneous beyond length scales of
$250 \hmpc$.

\end{abstract}


\section{Introduction}

The assumption that the Universe is statistically homogeneous and
isotropic on sufficiently large scales is fundamental to modern
cosmology. Our current understanding of the Universe comes from the
vast amount of cosmological observations which are hard to interpret
without relying on this assumption. Therefore it is important to
verify this assumptions using various observations.  There are a
multitude of evidences favouring isotropy such as the isotropy of the
CMBR \citep{penzias,smoot,fixsen}, isotropy in angular distributions
of radio sources \citep{wilson,blake}, isotropy in the X-ray
background \citep{peeb93,wu,scharf}, isotropy of Gamma-ray bursts
\citep{meegan,briggs}, isotropy in the distribution of galaxies
\citep{marinoni,alonso}, isotropy in the distribution of supernovae
\citep{gupta,lin} and isotropy in the distribution of neutral hydrogen
\citep{hazra}. But the local isotropy around us alone is not
sufficient to assure large-scale statistical homogeneity. One requires
to combine the local isotropy with the Copernican principle to infer
the large-scale statistical homogeneity of the Universe. The
Copernican principle states that we do not occupy a special location
in the Universe which itself requires validation. One can infer the
large-scale statistical homogeneity from local isotropy only when it
is assured around each and every point in the Universe. So it is not
straightforward to infer large-scale statistical homogeneity of the
Universe from the local isotropy.

A large number of studies
\citep{martinez94,borgani95,guzzo97,cappi,bharad99,pan2000,yadav,hogg,prakash,scrim,nadathur,pandey15,pandey16}
find that the galaxy distribution behaves like a fractal on small
scales but on large-scale the Universe is homogeneous. Most of these
studies claim to have found a transition to homogeneity on scales
$70-150 \, h^{-1} \rm {Mpc}$. Contrary to these claims a number of
studies \citep{pietronero,coleman92,amen,joyce,labini07, labini09,
  labini11} reported multi-fractal behaviour on different length
scales without any transition to homogeneity out to the scale of the
survey. The results from these studies clearly indicates that there is
no clear consensus in this issue yet. There would be a major paradigm
shift in cosmology if the assumption of cosmic homogeneity is ruled
out with high statistical significance by multiple data sets.

The most important implication of inhomogeneities comes from the
averaging problem in General Relativity through their effect on the
large scale dynamics known as backreaction mechanism. The backreaction
mechanism is known to cause a global cosmic acceleration without any
additional dark energy component \citep{buchert97, schwarz, kolb06,
  buchert08,ellis}.

The Sloan Digital Sky Survey (SDSS) \citep{york} is the largest and
finest galaxy redshift survey todate. The quasars are the brightest
class of objects known as Active Galactic Nuclei (AGN). The high
luminosities of quasars allow them to be detected out to larger
distances. The SDSS DR12 quasar catalogue provides us an unique
opportunity to test the assumption of cosmic homogeneity on the
largest accessible scale due to its enormous volume coverage. The
presence of large quasar groups (LQG) in the quasar distribution is
known for quite some time. \citet{clowes} identified a huge LQG with
characteristic size $\sim 500 \hmpc$ at $z\sim 1.3$ in the DR7 quasar
catalogue and claimed that this structure is incompatible with
large-scale homogeneity indicating possible violation of the
cosmological principle. A number of subsequent studies
\citep{nadathur,marinello} pointed out some flaws in interpreting LQGs
as structures. However if such structures really exist then they owe
an explanation. In the present study we test the large-scale
homogeneity in the SDSS DR12 catalogue using information theory based
methods \citep{pandey13,pandey16}. We do not address the LQGs
separately but the presence of any such structures in the quasar
distribution are clearly expected to boost the signal of inhomogeneity
up to noticeably larger length scales.

Throughout our work, we have used the flat $\Lambda$CDM cosmology with
$\Omega_m = 0.3, \; \Omega_{\Lambda} = 0.7 \mbox{ and } h = 1$.

A brief outline of the paper follows. In section 2 we describe the
method of analysis followed by a description of the data in section
3. We present the results and conclusions in section 4.

\begin{figure*}
\resizebox{7cm}{!}{\rotatebox{0}{\includegraphics{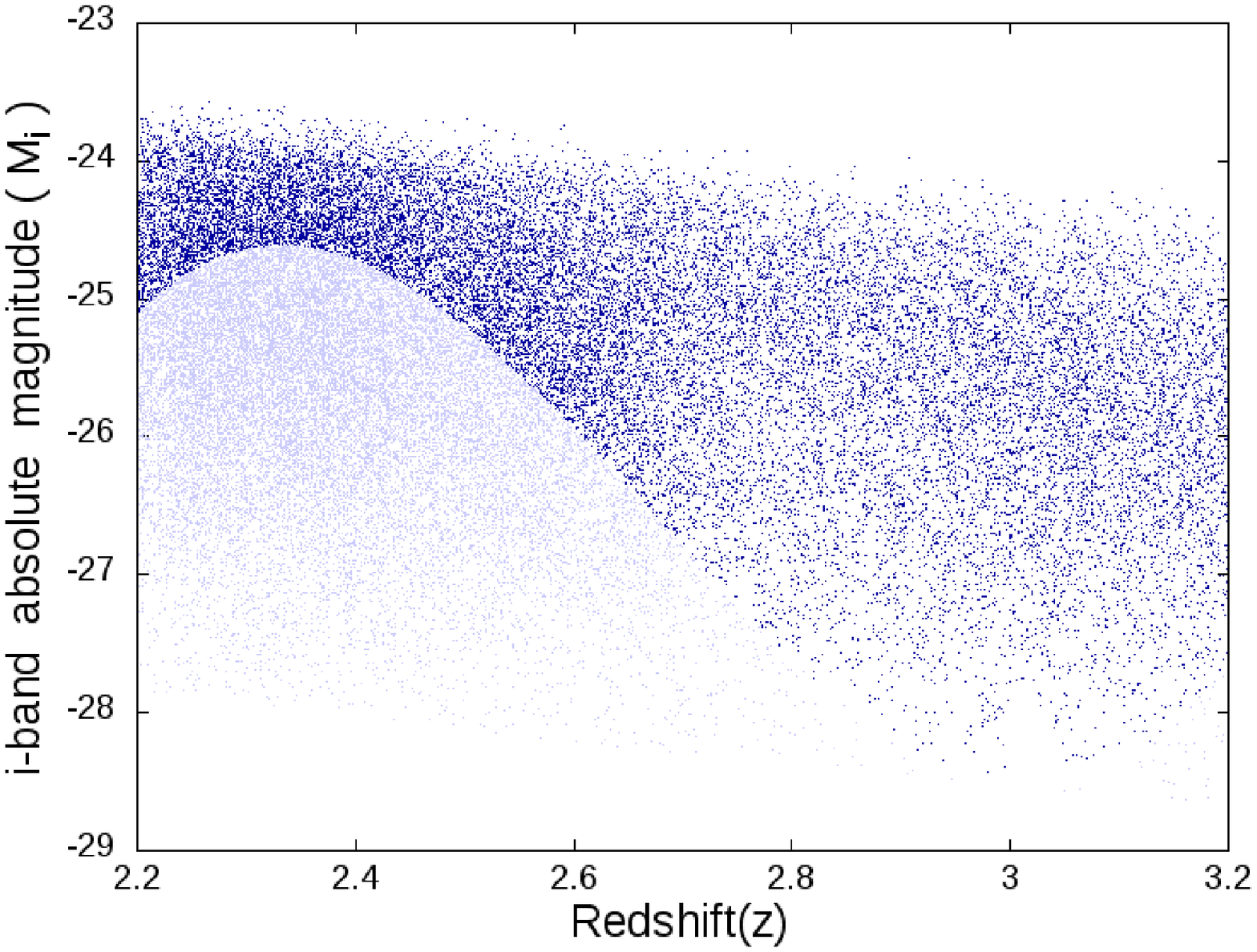}}}%
\resizebox{9cm}{!}{\rotatebox{0}{\includegraphics{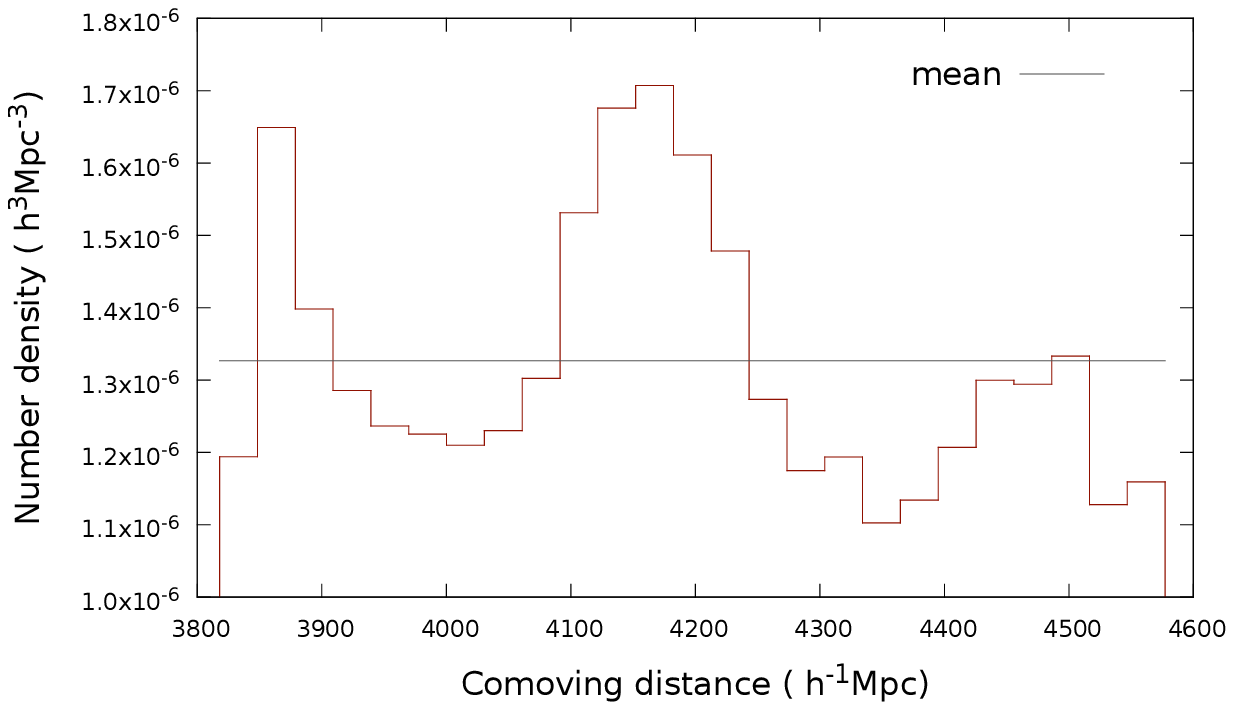}}}\\
\caption{The left panel shows the SDSS quasars in the
  redshift-absolute magnitude plane. The upper region and the lower
  region in this panel represent the selected and discarded quasars
  respectively. The right panel shows the comoving number density of
  quasars as a function of radial distance $r$. We compute the density
  in shells of uniform thickness $ 30.73 \hmpc$ in the radial direction.}
  \label{fig:qsoden}
\end{figure*}

\begin{figure*}
\resizebox{7cm}{!}{\rotatebox{0}{\includegraphics{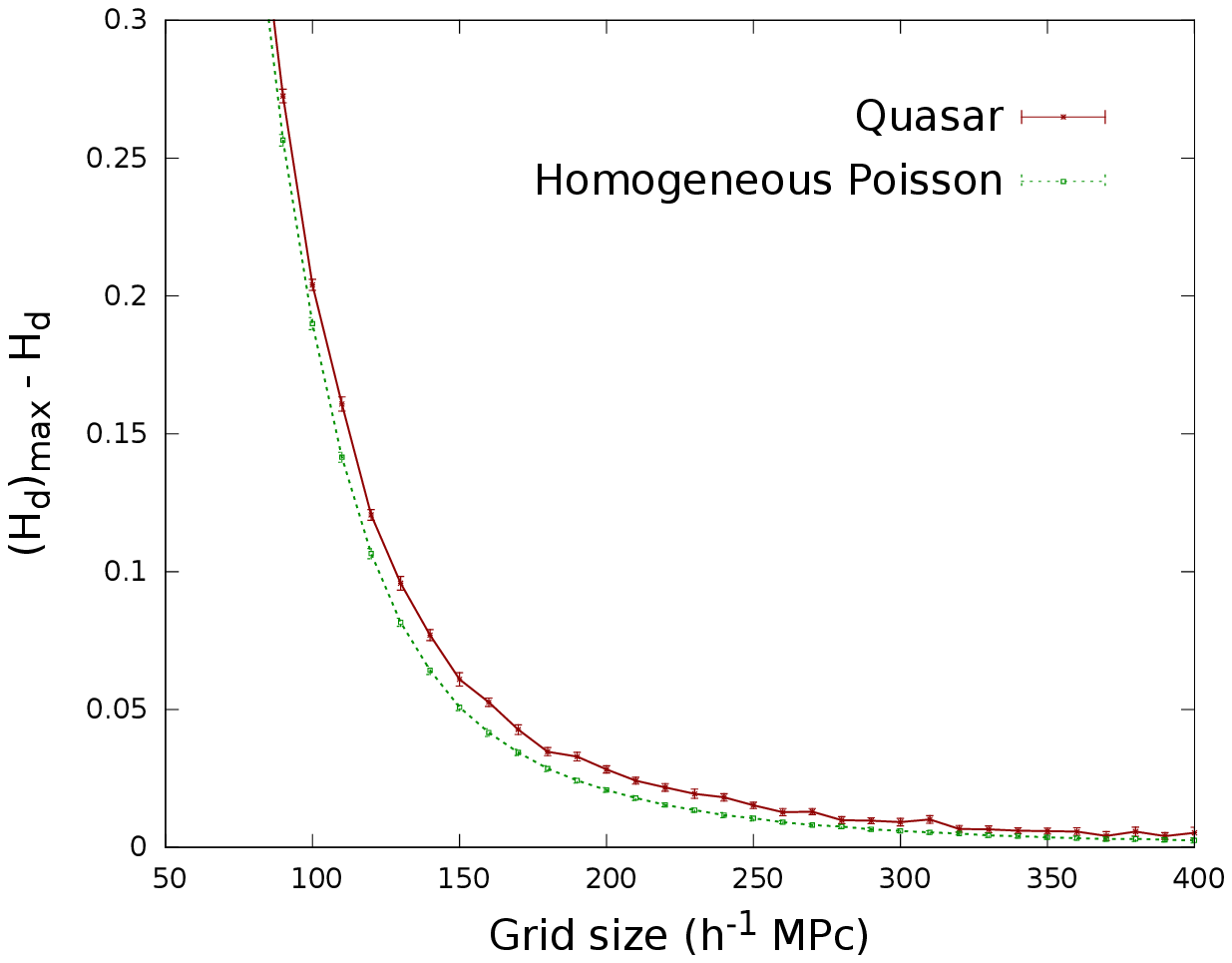}}}%
\resizebox{7cm}{!}{\rotatebox{0}{\includegraphics{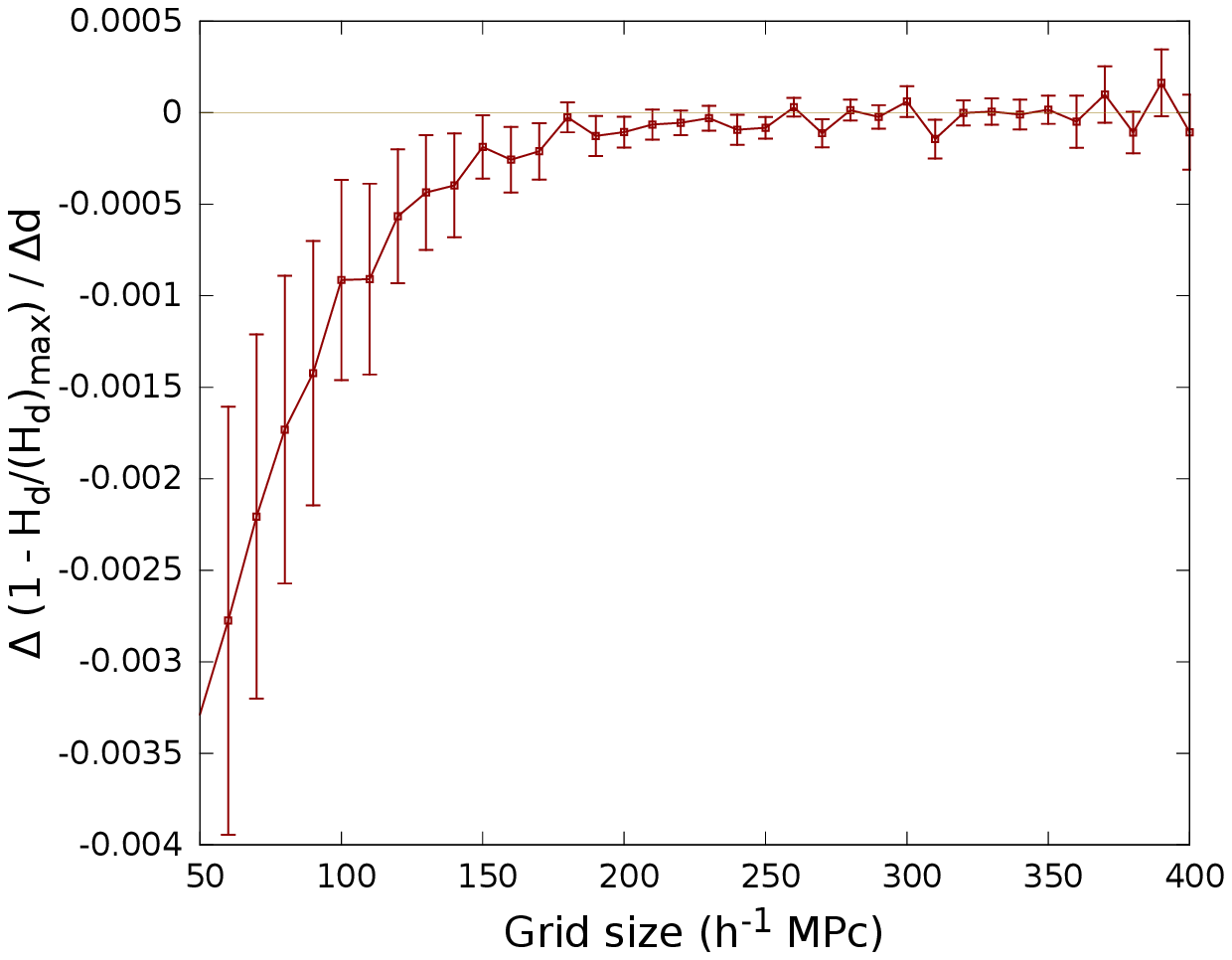}}}\\
\resizebox{7cm}{!}{\rotatebox{0}{\includegraphics{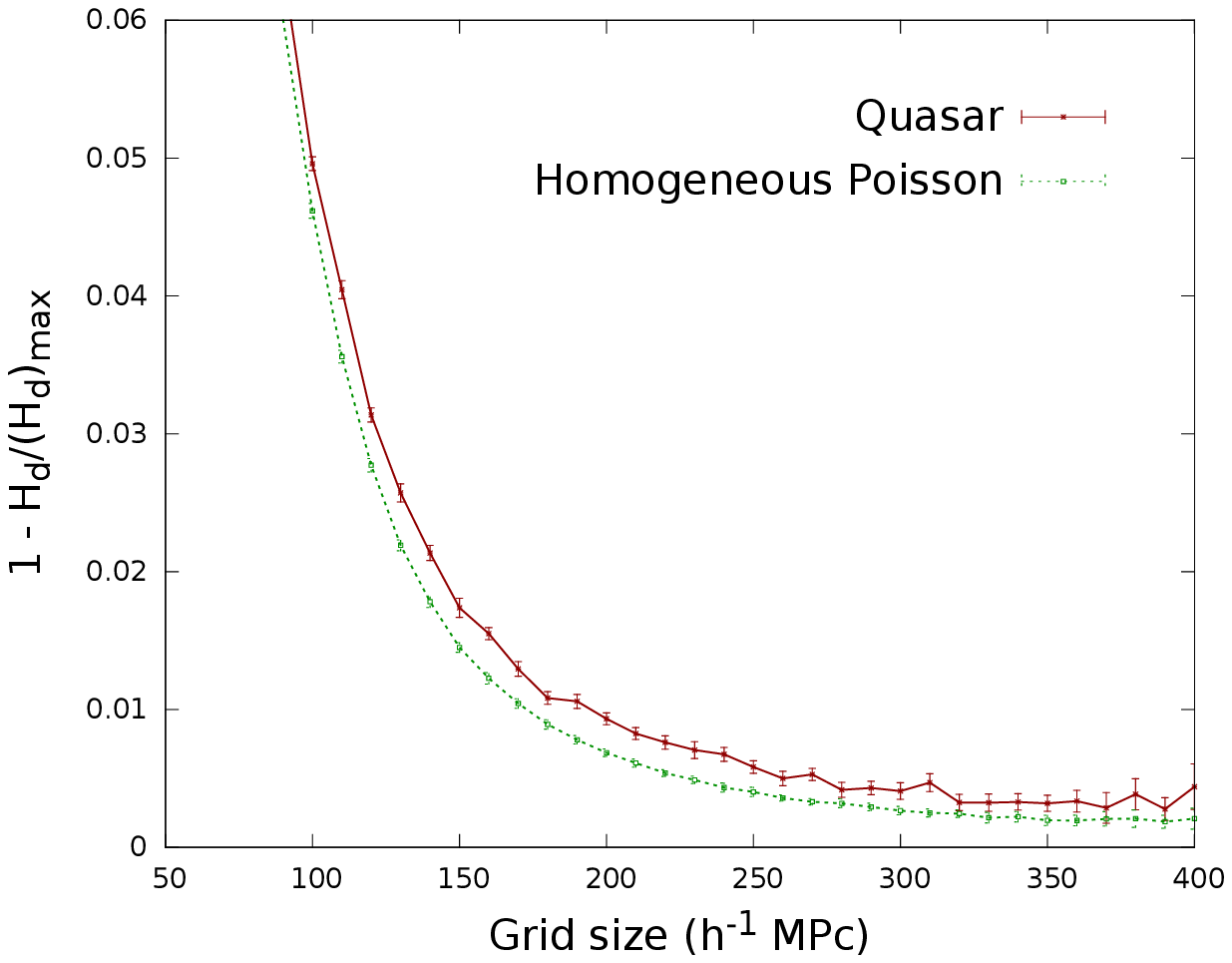}}}%
\resizebox{7cm}{!}{\rotatebox{0}{\includegraphics{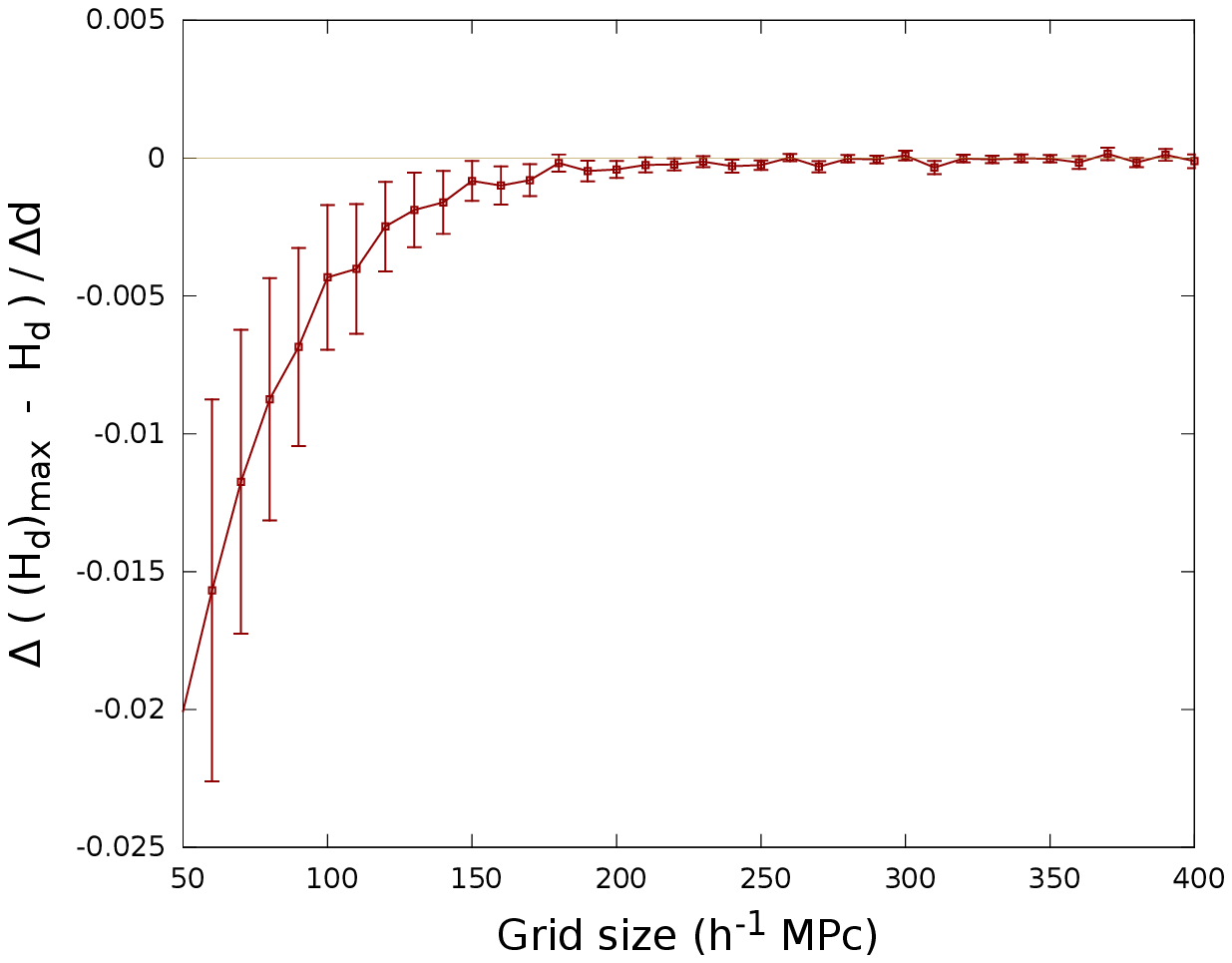}}}\\
\resizebox{7cm}{!}{\rotatebox{0}{\includegraphics{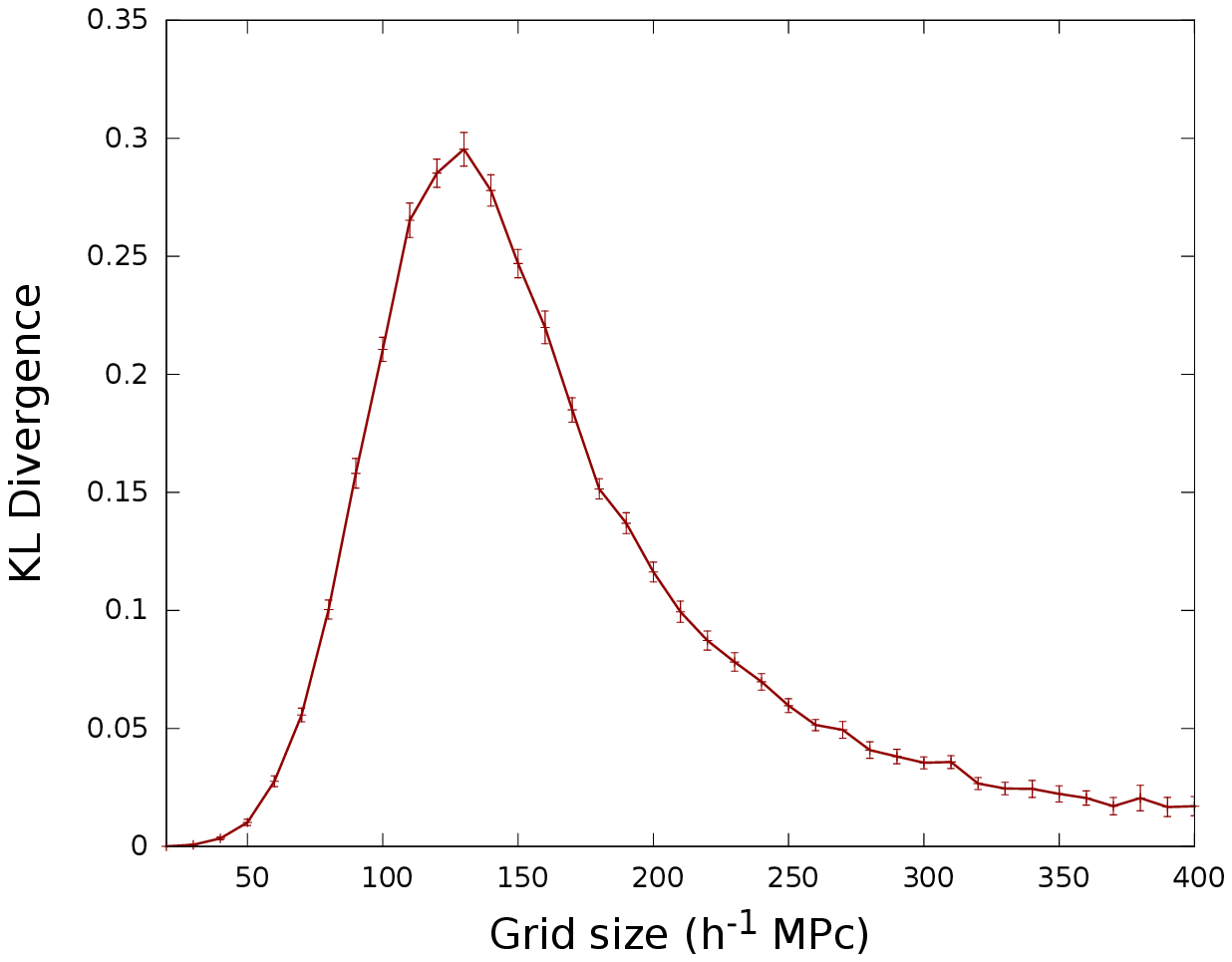}}}%
\resizebox{7cm}{!}{\rotatebox{0}{\includegraphics{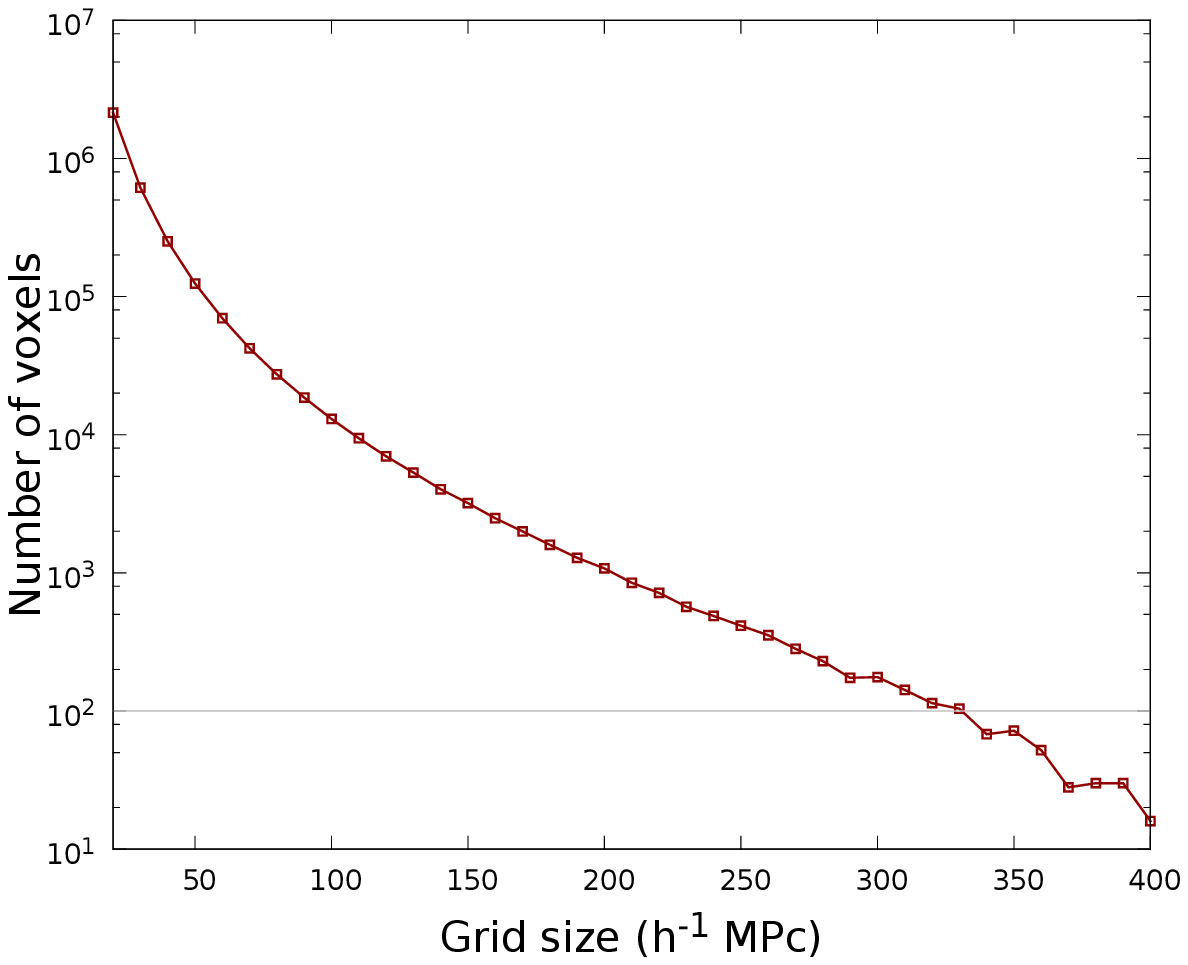}}}\\
\caption{The top left and middle left panels show
  $1-\frac{H_{d}}{(H_{d})_{max}}$ and $(H_{d})_{max}-H_{d}$ as a
  function of length scales respectively. The slopes of the respective
  quantities in the quasar sample are shown in the top right and
  middle right panels. The bottom left panel shows the KL divergence
  as a function of length scales and the bottom right panel shows the
  number of voxels available at each length scales. The errorbars for
  the quasar sample are obtained from $30$ bootstrap samples and the
  errorbars for the Poisson samples are obtained from $30$ Monte Carlo
  realizations of the same.}
  \label{fig:qsores}
\end{figure*}

\section{METHOD OF ANALYSIS}

Our analysis is based on a method proposed by \citep{pandey13} and its
extension \citep{pandey16} which used the Shannon entropy
\citep{shannon48} to study inhomogeneities in a 3D distribution. It
gives a measure of the average amount of information required to
describe a random variable. The Shannon entropy for a discrete random
variable $X$ with $n$ outcomes $\{x_{i}:i=1,....n\}$ is a measure of
uncertainty denoted by $H(X)$ defined as,

\begin{equation}
H(X) =  - \sum^{n}_{i=1} \, p(x_i) \, \log \, p(x_i)
\label{eq:shannon1}
\end{equation}

where $p(x)$ is the probability distribution of the random variable
$X$.  

We first embed the quasar distribution in a $d \, h^{-1}\, {\rm Mpc}
\times d \, h^{-1}\, {\rm Mpc} \times d \, h^{-1}\, {\rm Mpc}$ three
dimensional rectangular grid which divides the entire survey region
into a number of regular cubic voxels. We then identify the voxels
which lie partly or completely outside the survey region and discard
them to avoid any spurious effects from the boundary. Only the voxels
which are fully inside the survey region are retained for the
analysis. The grid size $d$ is varied within a suitable range and each
choice of $d$ result into a different number of voxels and quasars
within the survey region. Let $N_{d}$ be the number of voxels with
grid size $d$ and $n_{i}$ be the number of quasars inside the $i^{th}$
voxel. Now if we randomly pick up a quasar it can reside only in one
of the $N_{d}$ voxels i.e. there are $N_{d}$ possible outcomes as for
the position of this given quasar and the probability of finding the
aforementioned quasar in voxel $i$ is given by,
$f_{i,d}=\frac{n_{i}}{\sum^{N_{d}}_{i=1} \, n_{i}}$ with the
constraint $\sum^{N_{d}}_{i=1} \, f_{i,d}=1$. We denote the outcome of
the experiment with a random variable $X_{d}$ and the Shannon entropy
associated with the random variable $X_{d}$ can be written as,
\begin{eqnarray}
H_{d}& = &- \sum^{N_{d}}_{i=1} \, f_{i,d}\, \log\, f_{i,d} \nonumber\\ &=& 
\log(\sum^{N_{d}}_{i=1}n_{i}) - \frac {\sum^{N_{d}}_{i=1} \,
  n_i \, \log n_i}{\sum^{N_{d}}_{i=1} \, n_i}
\label{eq:shannon2}
\end{eqnarray}
Where the base of the logarithm is arbitrary and we choose it to be
$10$. 

We use three different measures of inhomogeneity based on the Shannon
entropy defined in \autoref{eq:shannon2} and describe them in the
following subsections.

\subsection{Relative Entropy and Entropy Deficit}

The probability of finding a quasar in the $i^{th}$ voxel $f_{i,d}$
will have the same value $\frac{1}{N_{d}}$ for all the voxels when
$n_{i}$ become the same for all of them. This is an ideal situation
when each of the $N_{d}$ voxels available contain exactly the same
number of quasars within them. This maximizes the Shannon entropy to
$(H_{d})_{max}=\log \, N_{d}$ for grid size $d$. We define the
relative Shannon entropy as the ratio of the entropy of a random
variable $X_{d}$ to the maximum possible entropy $(H_{d})_{max}$
associated with it. The relative Shannon entropy
$\frac{H_{d}}{(H_{d})_{max}}$ for any grid size $d$ quantifies the
degree of uncertainty in the knowledge of the random variable
$X_{d}$. Equivalently $1-\frac{H_{d}}{(H_{d})_{max}}$ quantifies the
residual information and can be treated as a measure of
inhomogeneity. The fact that quasars are not residing in any
particular voxel and rather are distributed across the available
voxels with different probabilities acts as the source of
information. If all the quasars would have been residing in one
particular voxel leaving the rest of them as empty then there would be
no uncertainty at all making $H_{d}=0$ or
$1-\frac{H_{d}}{(H_{d})_{max}}=1$. This fully determined hypothetical
situation corresponds to maximum inhomogeneity. On the other hand when
all the $N_{d}$ voxels are populated with equal probabilities it would
be most uncertain to decide which particular voxel a randomly picked
quasar belongs to. This maximizes the information entropy to
$H_{d}=\log \, N_{d}$ turning $1-\frac{H_{d}}{(H_{d})_{max}}=0$.

As $H_{d}=(H_{d})_{max}$ represents a homogeneous distribution we
define the entropy deficit $(H_{d})_{max}-H_{d}$ to quantify the
deviation of the distribution from uniformity. Clearly dividing the
entropy deficit by $(H_{d})_{max}$ provides the relative Shannon
entropy.

\subsection{Kullback-Leibler divergence}

Alternatively one can measure the inhomogeneity by using the
Kullback-Leibler (KL) divergence or information divergence
\citep{kullback,hosoya,li}. In information theory KL divergence is
used to measure the difference between two probability distributions
$p(x)$ and $q(x)$.\\
\begin{equation}
 D_{KL}(p|q)=\sum_{i} p(x_{i}) \, \log \frac{p(x_{i})}{q(x_{i})}
\label{eq:kld}
\end{equation}

Let $f_{Di}$ be the distribution corresponding to the data for which
homogeneity is to be tested and $f_{Ri}$ be the distribution for a
homogeneous and isotropic Poisson random distribution where both of
the distributions occupy same 3D volume with identical geometry and
are represented by same number of points.

The KL divergence between the actual and random data is then given
by,\\
\begin{equation}
 D_{KL}(D|R)=\frac{(\sum_{i} n_{Di}\,\log n_{Di}-\sum_{i} n_{Di}\,\log
   n_{Ri})}{\sum_{i} n_{Di}}-\log\frac{\sum_{i} n_{Di}}{\sum_{i} n_{Ri}}
\label{eq:kldiv}
\end{equation}

where $n_{Di}$ and $n_{Ri}$ are the counts in the $i^{th}$ voxel for
actual data and random data respectively. We use natural logarithm in
\autoref{eq:kldiv}.

\section{DATA}
\subsection{SDSS DR12 QUASAR SAMPLE}
We use the SDSS DR12 quasar catalog which includes $297301$ quasars.
The data is downloaded from the link
http://www.sdss.org/dr12/algorithms/boss-dr12-quasar-catalog. The
quasar target selection is described by \citet{rossq} and the data is
described by \citet{paris}. We first exclude the objects with a
ZWARNING value that calls into question the accuracy of their redshift
determination. We apply the criteria ZWARNING $=0$ \citep{bolton} and
then set UNIFORM $\geq 1$ to identify a homogeneously selected sample
of quasars. This provides the CORE targets selected using XDQSO
technique after chunk 12 \citep{bovy} and also those which would have
been selected by XDQSO if it had been the core algorithm prior to
chunk 12. We apply further cuts in g-band PSF magnitude $g\leq22.0$
and r-band PSF magnitude $\leq21.85$ \citep{rossq} to get $117882$
quasars with XDQSO probability greater than $0.424$. We then construct
our quasar sample covering a contiguous region of right ascension
$150^{\circ} \leq \alpha \leq 240^{\circ}$ and declination $0^{\circ}
\leq \delta \leq 60^{\circ}$ in the redshift range $2.2 \leq z \leq
3.2$. The resulting quasar sample does not have an uniform number
density in the entire redshift range. We find that constructing a
volume limited sample of quasars from this data results into a sample
with a very poor number density. So to construct a quasar sample we
use a cut $M_{i} \geq M_{lim}(z)$ in the i-band absolute magnitude
which varies with redshift. We describe $M_{lim}(z)$ with a polynomial
as $M_{lim}(z)= a z^{3} + b z^{2} + c z + d$ where $a,b,c$ and $d$ are
the coefficients to be determined. We constrain these coefficients so
as to produce a quasar sample with a near uniform comoving number
density. We find that the coefficients
$a=24.206,b=-194.325,c=511.413,d=-467.422$ produces a quasar sample
consisting of $24213$ quasar which has less than $30\%$ variation
(\autoref{fig:qsoden}) in number density around the mean in the entire
redshift range. The sample has a volume of $1.83 \times 10^{10}$
${(\hmpc)}^3$ with a linear extent of $759.32 \hmpc$ in the radial
direction. The mean number density of the quasar sample is $1.326
\times 10^{-6} \, h^{3}\, {\rm Mpc}^{-3}$.

\subsection{RANDOM SAMPLES}

We construct $30$ mock random samples from homogeneous and isotropic
3D Poisson point processes. The mock random samples contain exactly
the same number of points as there are quasars in our sample and are
distributed within a region which have the same geometry as the quasar
sample. 

\section{RESULTS AND CONCLUSIONS}

We present our results in \autoref{fig:qsores}. In top left and middle
left panel of the \autoref{fig:qsores} we show respectively the
variation of $1-\frac{H_{d}}{(H_{d})_{max}}$ and $(H_{d})_{max}-H_{d}$
with increasing grid sizes for the quasar sample and its random mock
counterparts. Both the quasar sample and the random samples have a
mean inter-particle separation of $91 \hmpc$ and they are hardly
distinguishable below this length scale but as the grid size increases
the differences become evident. In both of the plots we find that the
quasar sample has a higher information content than the Poisson
samples throughout the entire length scale ranges due to the
gravitational clustering. However the differences diminish with
increasing length scales and at $\sim 250-300 \hmpc$ the results for
the Poisson samples lies within $1-\sigma$ errorbars of the same for
the quasar sample. The residual inhomogeneities beyond $250 \hmpc$ are
consistent with what one would expect for a homogeneous Poisson point
process. We show the rates of change of
$1-\frac{H_{d}}{(H_{d})_{max}}$ and $(H_{d})_{max}-H_{d}$ in the
quasar sample with increasing length scales in the top right and
middle right panels of \autoref{fig:qsores} respectively. We find that
the rates of change for both of the measures in the quasar sample
diminish nearly to zero with tiny errorbars at $\sim 250 \hmpc$. In
the bottom left panel of \autoref{fig:qsores} we show the KL
divergence measure as a function of length scales in the quasar
sample. The KL divergence also indicates that the inhomogeneities
diminish with increasing length scales finally reaching a plateau at
$\sim 250-300 \hmpc$. However it is worth mentioning here that though
the entropy is sensitive to the higher order moments of a distribution
it may not capture the signatures of the full hierarchy of correlation
functions. A Minkowski Functional analysis of SDSS LRGs by
\citet{wiegand} find significant deviations from the $\Lambda$CDM
mock catalogues on scales of $500 \hmpc$.

It has been suggested that the quasars inhabit dark matter halos of
constant mass $\sim 2 \times 10^{12} h^{-1} M_{\odot}$ from redshifts
$z \sim 2.5$ (the peak of quasar activity) to $z\sim 0$ and their
large scale linear bias evolves from $b=3$ at $z\sim 2.2$ to $b=1.38$
at $z\sim 0.5$ \citep{shen, ross, geach}. Our quasar sample extends
from $z=2.2$ to $z=3.2$ for which we expect a large scale linear bias
of $\gtrsim 3$. As the quasars inhabit rarer high density peaks one
would expect the quasar sample to be homogeneous on even larger scale
than the SDSS main galaxy sample and the LRG sample. Our result is
consistent with our earlier studies on the SDSS main galaxy sample
\citep{pandey15} and the LRG sample \citep{pandey16} for which we find
a transition scale to homogeneity at $\sim 150 \hmpc$.

It may be noted here that there are $414$ independent voxels (bottom
right panel of \autoref{fig:qsores}) at grid size of $250 \hmpc$ and
each voxel is expected to host $\sim 21$ quasars provided the
distribution is homogeneous beyond this length scale. This number is
certainly very small due to the small number density of the quasar
sample. Further we can not have access to spatial hypersurface of
constant time. So any analysis of homogeneity on large scales would
unavoidably incorporate some signatures of the time evolution. Despite
these difficulties it is interesting to note the degree of homogeneity
in the quasar sample beyond a length scale of $250 \hmpc$. Last but
not the least we prepare $4$ different quasar samples with different
set of values for the coefficients $(a,b,c,d)$ used to define the
limiting magnitudes at different redshifts. Irrespective of our choice
we find the same transition scale to homogeneity in each of these
quasar samples. We finally conclude that the SDSS quasar distribution
is homogeneous beyond $250 \hmpc$, for the information theoretic
measures employed in this paper.

\section{ACKNOWLEDGEMENT}
The authors would like to thank the SDSS team for making the data
publicly available. B.P. would like to acknowledge IUCAA, Pune and
CTS, IIT Kharagpur for the use of its facilities for the present work.

% Don't change these lines
\bsp	% typesetting comment
\label{lastpage}
\end{document}